# Exploring the Interconnectedness of Cryptocurrencies using Correlation Networks

Andrew Burnie
UCL Computer Science Doctoral Student at The Alan Turing Institute
aburnie@turing.ac.uk

## Abstract

Correlation networks were used to detect characteristics which, although fixed over time, have an important influence on the evolution of prices over time. Potentially important features were identified using the websites and whitepapers of cryptocurrencies with the largest userbases. These were assessed using two datasets to enhance robustness: one with fourteen cryptocurrencies beginning from 9 November 2017, and a subset with nine cryptocurrencies starting 9 September 2016, both ending 6 March 2018. Separately analysing the subset of cryptocurrencies raised the number of data points from 115 to 537, and improved robustness to changes in relationships over time. Excluding USD Tether, the results showed a positive association between different cryptocurrencies that was statistically significant. Robust, strong positive associations were observed for six cryptocurrencies where one was a fork of the other; Bitcoin / Bitcoin Cash was an exception. There was evidence for the existence of a group of cryptocurrencies particularly associated with Cardano, and a separate group correlated with Ethereum. The data was not consistent with a token's functionality or creation mechanism being the dominant determinants of the evolution of prices over time but did suggest that factors other than speculation contributed to the price.

**Keywords**: Correlation Networks; Interconnectedness; Contagion; Speculation

# 1. Introduction

The year 2017 saw the start of a rapid diversification in cryptocurrencies. Over 800 new cryptocurrencies were launched (BitInfoCharts, 2018), coinciding with a decline in Bitcoin's dominance of the marketplace. Whilst Bitcoin represented over 85% of total market capitalisation at the beginning of 2017, by the end, this had fallen to less than 40% (CoinMarketCap, 2018).

This diversity of cryptocurrencies reflects a variety of potential uses. These range from the specific, such as providing a new type of money (e.g. Bitcoin Cash) or the underpinnings of a decentralised storage network (e.g. Filecoin), to the generic, providing a tool for application development (e.g. EOS and Qtum).

This diversity also, however, complicates understanding cryptocurrencies from a valuation perspective. It is unclear which of a given cryptocurrency's heterogenous characteristics or what in the external context of the cryptocurrency should be focussed on in understanding whether the associated token is under- or over-valued. At the macro-level, some thought leaders have even suggested that cryptocurrencies do not provide participants with a genuine source of value (Bercetche, 2017; Imbert, 2017) and so are all overvalued. This would suggest that investors are driven by just an irrational response to rising prices – i.e. by speculation (Shiller, 2003). A lack of clarity as to what drives individual cryptocurrency valuations prevents assessments of the true impact of speculation.

Previously, the literature has tended to focus on trying to understand which features are important from the perspective of whether a change in a particular feature is associated with a subsequent change in the cryptocurrency price. Typically, this has involved comparing a variety of candidate value drivers using an econometric model (Bouoiyour & Selmi, 2015; Ciaian, Rajcaniova, & Kancs, 2016; Garcia & Schweitzer, 2015; Garcia, Tessone, Mavrodiev, & Perony, 2014; Kristoufek, 2013; Wijk, 2013) or wavelet coherence analysis (Kristoufek, 2015). Consideration of cryptocurrencies other than Bitcoin has been limited to a few examples, such as Ripple and Ethereum (Kim et al., 2016).

Such an approach is limited in being unable to detect characteristics which, although fixed over time, have an important influence on the evolution of prices over time. A possible example would be Ethereum's smart contract technology. Ethereum has had smart contracts from its launch; this characteristic is time-invariant. Whether smart contracts have altered the evolution of the price of ether over time thus cannot be determined from ether's price data alone. The applied technique needs to include a comparison with other cryptocurrencies.

This paper aims to compare cryptocurrencies to infer important time-invariant characteristics that shape the development of prices over time. First theoretical suggestions are made as to what these characteristics might be; the implied groupings are then empirically tested using correlation network diagrams. If two cryptocurrencies share common important characteristics, this is likely to manifest in a strong positive correlation.

Examining the correlations between cryptocurrencies further helps us to understand the extent to which results found for one cryptocurrency (particularly Bitcoin) can be generalised to others. It will also help to reveal which cryptocurrencies are particularly sensitive to contagion effects where declines in one cryptocurrency affects others. Finally, it will help investors designing cryptocurrency portfolios to invest in uncorrelated cryptocurrencies, thus diversifying their overall risk.

This paper improves on previous, related work in: comparing a variety of different cryptocurrencies with each other, rather than just comparing Bitcoin against others (Ciaian, Rajcaniova, & Kancs, 2018; Gandal & Halaburda, 2016); including cryptocurrencies with smart contract functionality (Gandal & Halaburda, 2016; Osterrieder, Lorenz, & Strika, 2017); and in evaluating more broadly whether cryptocurrencies are associated with each other rather than specifically testing for a linear relationship (Ciaian et al., 2018). This paper also explicitly analyses whether there is evidence to support the existence of value-relevant, time-invariant features, and evidence to support or reject potential candidates.

# 2. Methodology

## 2.1 Rationale for Cryptocurrencies Selected

Cryptocurrencies were selected that had a large userbase relative to other cryptocurrencies.

If a cryptocurrency has a small userbase, then the number of buyers and sellers at any given point in time is likely to be smaller. Hence, a buyer of the cryptocurrency is likely to need to raise prices further to induce sufficient supply; a seller will need to lower prices further to encourage sufficient demand. This suggests that the prices of smaller cryptocurrencies will be more volatile, and that this volatility will be driven by random noise.

A smaller cryptocurrency is also likely to be listed on fewer exchanges. With fewer tokens being bought or sold at any one time, the incentive for exchanges to list a given cryptocurrency is less. Smaller cryptocurrencies will thus be more prone to the difficulties faced by specific exchanges, such as an exchange being hacked (Rosic, 2017), trading ceasing due to technological reasons (Greene, 2017), or bankruptcy (Meyer, 2017).

Hence, cryptocurrencies with smaller userbases were avoided because their price series were more likely to be driven by idiosyncratic noise. This would weaken the cryptocurrency's correlation with others in the dataset, raising the risk of concluding that pairs of cryptocurrencies did not share important common characteristics when, in truth, they did.

To ensure the same price data was used throughout, data was sourced from coingecko.com to determine the cryptocurrencies with the largest userbases on 6 March 2018. Two financial measures of the userbase were available: market capitalisation and liquidity.

Market capitalisation measures the price of the token multiplied by the available supply and so directly measures the amount held in each cryptocurrency (CoinGecko, 2018). A limitation to existing metrics of market capitalisation is how to account for inaccessible tokens resulting from owners losing access to their wallets or hoarding (Torpey, 2016). Such scenarios could lead to market capitalisation giving a misleading impression of the amount invested in a given cryptocurrency.

We thus also considered liquidity, which here measures the trading activity across exchanges (CoinGecko, 2018). The fewer tokens available for sale, the lower the likely transaction volume for a given market capitalisation.

Any cryptocurrency in the top ten by market capitalisation or liquidity was analysed.

## 2.2 Questions used to Demarcate Cryptocurrencies

An approach for consideration would be to derive groupings using the frameworks proposed by regulators. Regulators have often been reluctant in providing specific guidance (Financial Conduct Authority, 2017; Monetary Authority of Singapore, 2017); exceptions being the United States Securities and Exchange Commission (SEC), the Commodity Futures Trading Commission (CFTC) and the Swiss Financial Market Supervisory Authority (FINMA).

The SEC recommends differentiating tokens according to whether or not they are securities, involving the 'reasonable expectation of profits to be derived from the entrepreneurial or managerial efforts of others' (United States Securities and Exchange Commission, 2017). FINMA corroborates with such an approach, splitting security tokens further between utility and asset tokens, and describing payment tokens as being subject to anti-money laundering regulation but not security regulation (Swiss Financial Market Supervisory Authority (FINMA), 2018). Such a split between securities and non-securities is difficult to apply in practice (Bennington, 2017) and complicated by the CFTC instead advocating differentiation between commodities and derivatives, both subject to their oversight (Higgins, 2017).

An underlying theme to the above regulator discussions has been a focus on the function of the token (Burnie, Henderson, & Burnie, 2018), where the token is the digital representation of value in the cryptocurrency system. This paper will broaden the characteristics considered to include the different stages of a token's life-cycle (Figure 1). Each stage has a relevance to either the supply or demand of tokens, and so could potentially influence how prices evolve over time.

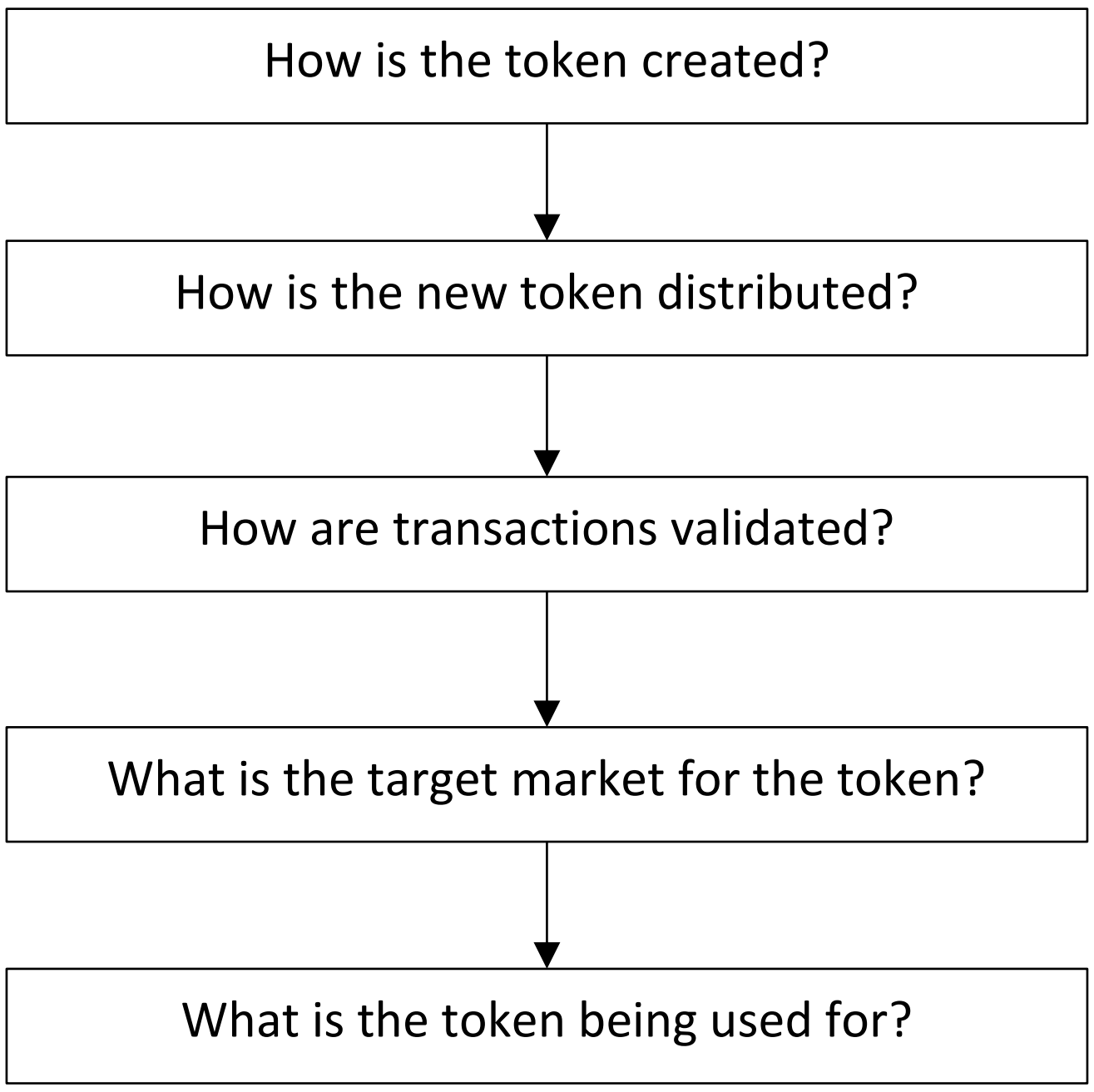

*Figure 1: Questions relevant at different stages of the lifespan of a token*

The questions in Figure 1 were applied to the subset of cryptocurrencies selected (Section 2.1) using their websites and whitepapers to provide answers. This identified common characteristics which could potentially drive a strong correlation between similar cryptocurrencies.

The resulting groups of cryptocurrencies were compared against two alternate scenarios in explaining the price series. The first is that all cryptocurrencies share common fixed characteristics that drive the evolution of prices over time. The second is that all cryptocurrencies are unrelated and distinct.

### 2.3 Selecting the Dataset

Daily pricing data (in USD) was gathered from coingecko.com on 6 March 2018 for each cryptocurrency considered. This source was preferred because it covered a wide variety of cryptocurrencies and enabled the downloading of data in CSV format. This source was limited in having missing data for certain dates: 22 February 2018 had missing data for 11 of the cryptocurrencies; NEO had missing data for 8-10 August 2017. The dataset began from 28 April 2013 at the earliest.

Rather than comparing the raw price series, the daily percentage change in price for each cryptocurrency was calculated. This provided a closer proxy to the returns an investor would have received if they had held a particular cryptocurrency on a certain day. As this calculation involved first differencing, it was also more robust should there be nonstationarity problems in the dataset (Stock & Watson, 2012).

As different cryptocurrencies were launched in different years, data availability varied. We thus created two datasets. In the first dataset, we considered all the cryptocurrencies, which required beginning the time series from 9 November 2017. In the second dataset, we considered a subset of cryptocurrencies where there was more data, enabling us to begin our time series from 9 September 2016. Considering different time periods ensured greater robustness to the instability in correlation values over time (Gandal & Halaburda, 2016; Osterrieder et al., 2017).

## 2.4 Applying Correlation Networks

### 2.4.1 The Correlation Metric

If two cryptocurrencies' daily returns series are influenced by the same common characteristics, then these returns should be strongly associated. To test for such an association, a correlation metric is required.

Particularly popular correlation metrics are: Pearson's product moment correlation coefficient (PMCC); Spearman's rho (SR); and Kendall's tau (KT) (Mari & Kotz, 2001). Applying the PMCC assumes that cryptocurrency returns follow normal distributions (Xu, Hou, Hung, & Zou, 2013), which previous research has suggested to be an unreasonable assumption (Chan, Chu, Nadarajah, & Osterrieder, 2017; Osterrieder et al., 2017). The PMCC is further restricted in measuring linear relationships (Xu et al., 2013).

There is evidence to support SR as being a more accurate measure of the association between variables (in terms of Mean Squared Error) when the sample size is small and true population correlation is weak. In contrast, KT is supported as being more accurate when the true population correlation is strong and/or in large sample cases (Xu et al., 2013). Osterrieder, Lorenz, & Strika (2017) found that the correlations between cryptocurrencies were typically small (except for between Bitcoin and Litecoin) and the dataset size is limited (Section 2.3), thus this paper primarily uses the SR methodology, with KT being used to check for robustness.

The formula for the SR between the series x and y is as follows:

$$SR(x, y) = \frac{Cov\left(Rank_x, Rank_y\right)}{\sigma_{Rank_x}\sigma_{Rank_y}}$$

Where $Rank_i$ represents the ranks for the different values of variable i. The covariance between the ranked versions of x and y is normalised by dividing by the standard deviations of the ranked versions of x and y.

Robustness was checked through the use of KT; the formula is as follows:

$$KT(x,y) = \frac{n_{concordant} - n_{discordant}}{\sqrt{n_x} \times \sqrt{n_y}}$$

Where:

- $n_{concordant}$ is a count of the number of pairs of values where the ordering in series x matches that in y;
- $n_{discordant}$ is the number of pairs where the ordering does not match;
- pairs of tied values are ignored in the above counts;
- $n_i$ is the total number of pairs of datapoints that are not tied in series i.

The above formula relates to Kendall's Tau-b (Kendall, 1945). If there are no ties, then $n_x = n_y$ = the total number of possible pairs. The above equation then becomes equivalent to the following (Kendall, 1938):

$$KT(x,y) = \frac{n_{concordant} - n_{discordant}}{\frac{n(n-1)}{2}}$$

### 2.4.2 Correlation Networks

A network consists of circular nodes connected by lines called edges. In this paper's correlation network, the nodes represent the daily returns for different cryptocurrencies whilst each edge has a weight that is the correlation between the linked cryptocurrencies' returns. Diagrammatically, the stronger the association between two cryptocurrencies' returns, the wider the line connecting their nodes (Epskamp, Cramer, Waldorp, Schmittmann, & Borsboom, 2012).

To aid interpretability, the nodes were arranged such that more correlated cryptocurrencies are placed closer together. This cannot always be perfectly achieved in a two-dimensional space (Epskamp et al., 2012), so instead an approximate force-embedded algorithm approach was applied (Fruchterman & Reingold, 1991).

The correlation networks were refined to aid interpretability. This involved using a threshold value, below which links in the correlation network were removed. To improve robustness, the threshold value was selected in a variety of ways. In the first approach, all correlation values were arranged from largest to smallest and cut-offs placed where there was a jump in correlation value. In the second approach, only the top ten correlation values were depicted. In the third approach, the threshold value was increased in 0.01 increments until the cryptocurrencies were split into at least two separate groups.

The correlation networks were compared with the proposed classifications (Section 2.2) to determine the extent to which they were supported or refuted.

The networks were initially created using SR and robustness assessed using KT. When checking for robustness, a threshold was not placed where there was a jump in correlation value, because this thresholding approach was sensitive to the units of the correlation measure.

#### 2.4.3 Significance Tests

Two-sided significance tests were performed to evaluate if there was sufficient evidence to reject a null hypothesis of no association between the returns of different cryptocurrencies. This was performed using the SR measure of correlation, with KT being used to check for robustness.

#### 2.4.4 Software

Correlation networks were applied using the programming language R. The correlation matrices and tests were implemented using base R functions (cor and cor.test respectively), which did not require the installation of additional packages. The correlation network was implemented using the package qgraph (Epskamp et al., 2012), which was specifically designed for this process.

## 3. Results

### 3.1 Cryptocurrencies Considered

The top ten cryptocurrencies in terms of market capitalisation or liquidity were determined on 6 March 2018. The cryptocurrencies selected were: Bitcoin, Litecoin, Ethereum, Ethereum Classic, Monero, NEO, Bitcoin Cash, Tron, Cardano, Qtum, Ripple, EOS, Stellar and USD Tether.

These cryptocurrencies were also restricted to a subset with data available from 9 September 2016 rather than 9 November 2017: Bitcoin, Litecoin, Ripple, Monero, USD Tether, Ethereum, Ethereum Classic, Stellar and NEO.

### 3.2 Proposed Cryptocurrency Groups

Figures 2 and 3 provide potential cryptocurrency groups derived from the questions raised in Figure 1 about the different characteristics of the cryptocurrency token. These characteristics could be subject to change, particularly as five (Cardano, Bitcoin Cash, EOS, Qtum and Tron) of the cryptocurrencies were launched as recently as 2017.

**Token Supply and Transactions**

**Token Creation**

**Fixed**
- NEO
- Tron
- Cardano
- Qtum
- Ripple

**Rise up to Cap**
- Bitcoin
- Litecoin
- Ethereum Classic
- Monero
- Bitcoin Cash

**Rise Indefinitely**
- Ethereum
- Stellar
- EOS

**Varies to maintain peg**
- Tether

**Token Distribution / Validation**

**Proof-Of-Work**
- Bitcoin
- Litecoin
- Ethereum
- Ethereum Classic
- Monero
- Bitcoin Cash

Run on top of Proof-Of-Work systems:
- Tron (on top of Ethereum)
- Tether (on top of Bitcoin)

**Validators selected**
- Ripple

**Voting**
- NEO
- EOS
- Stellar (new token distribution involves voting; verification through Federated Byzantine Agreement)

**Proof-Of-Stake**
- Cardano
- Qtum

*Figure 2: Grouping cryptocurrencies according to similarities in supply characteristics and how transactions are performed.*

| Token Demand |
|---|

| Target Market | |
|---|---|
| **Generic**<br>• Bitcoin<br>• Litecoin<br>• Ethereum<br>• Ethereum Classic<br>• Monero (albeit emphasis on privacy)<br>• NEO<br>• Bitcoin Cash<br>• Tether | **Business-Orientated**<br>• Cardano<br>• Ripple<br>• EOS<br>• Stellar<br>• Qtum<br><br>**Content Creators on Internet**<br>• Tron |

| Token Function | | |
|---|---|---|
| **Transaction**<br>• Litecoin<br>• Monero<br>• Bitcoin Cash<br>• Ripple<br>• Stellar | **Hybrid**<br>• Bitcoin<br>• Ethereum<br>• Cardano | **Applications**<br>• NEO<br>• Tron<br>• Qtum<br>• EOS<br>• Ethereum Classic |

*Figure 3: Grouping cryptocurrencies according to similarities in demand characteristics.*

See Appendix for a list of links to the different websites and white papers used to inform the above groups.

Similarities in token supply, distribution and validation could be determined objectively, although the precise mechanisms used were sometimes tentative. For example, Ethereum plans to switch to Proof-Of-Stake validation (Ethereum, 2014/2018).

Examining the whitepapers and websites of different cryptocurrencies revealed a split between cryptocurrencies targeting business use and cryptocurrencies seeking to be used more generally. Tron's stated market was neither, instead targeting content creators on the Internet.

Business-orientated cryptocurrencies were defined to be those cryptocurrencies that were explicit in seeking specifically commercial applications of their technology, for payments (Ripple, Stellar), for developing applications (Qtum, EOS) or for both (Cardano). These cryptocurrencies include discussions on the optimality of their system for business use in their whitepapers or on their websites (see Appendix).

Generic cryptocurrencies were defined as cryptocurrencies that targeted a broad business and non-business audience. This did not preclude some optimisation for business use. For example, NEO's whitepaper advocates its suitability for businesses, but also aims to create a smart economy, suggesting an intention to be used as much by individuals as by businesses (NEO, 2017). Similarly, Ethereum's whitepaper discusses business applications as well as non-business uses such as online voting (Ethereum, 2014/2018).

In examining these resources, it was further found to be convenient to define three types of token functionality:

- **Transaction:** tokens designed and used primarily for transacting value.
- **Applications:** tokens designed to enable the development of applications.
- **Hybrid:** cryptocurrencies that carry both types of functionality, and where evidence was found to support both types of functionality being in use.

Differentiating between these categories was complicated by cryptocurrencies often being launched for one purpose and then evolving to be used for another. Ethereum was originally launched to enable decentralised applications (Ethereum, 2014/2018), yet has evolved to be popular among merchants (finder, 2017). Bitcoin was launched for transacting value (Nakamoto, 2008), yet the Omni Layer Protocol has given Bitcoin functionality to launch applications (Omni Team, 2017) such as Tether (Tether Ltd., 2016).

## 3.3 Correlation Results

In the results, the following abbreviations were used:

| CRYPTOCURRENCY | ABBREVIATION |
|---|---|
| **Bitcoin** | btc |
| **Litecoin** | ltc |
| **Ethereum** | eth |
| **Ethereum Classic** | etc |
| **Monero** | xmr |
| **NEO** | neo |
| **Bitcoin Cash** | bch |
| **Tron** | trx |
| **Cardano** | ada |
| **Qtum** | qtm |
| **Ripple** | xrp |
| **EOS** | eos |
| **USD tether** | usdt |
| **Stellar** | xlm |

*Table 1: Abbreviations used for cryptocurrencies throughout the results section*

In the correlation diagrams, dashed lines indicate a negative correlation value – one value tended to decline as the other value rose; full lines indicate a positive correlation value – one value tended to rise with the other. Wider and darker lines are indicative of a higher correlation value. Cryptocurrencies with a higher correlation tend to be placed closer together.

### 3.3.1 Considering All Cryptocurrencies

#### *3.3.1.1 Correlation Network Diagrams*

**a) All correlations**

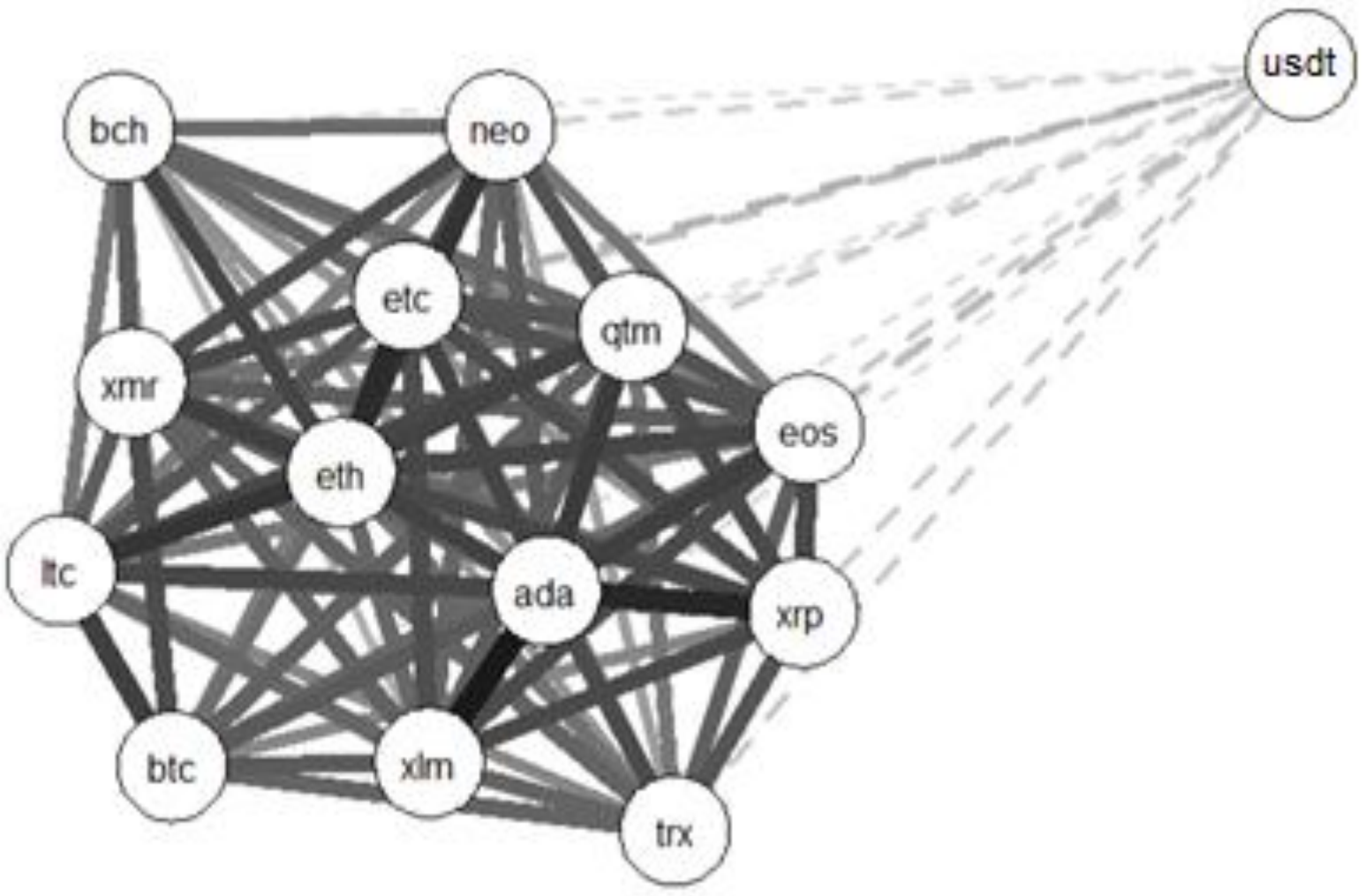

**b) Threshold where correlations jump in value (at 0.55)**

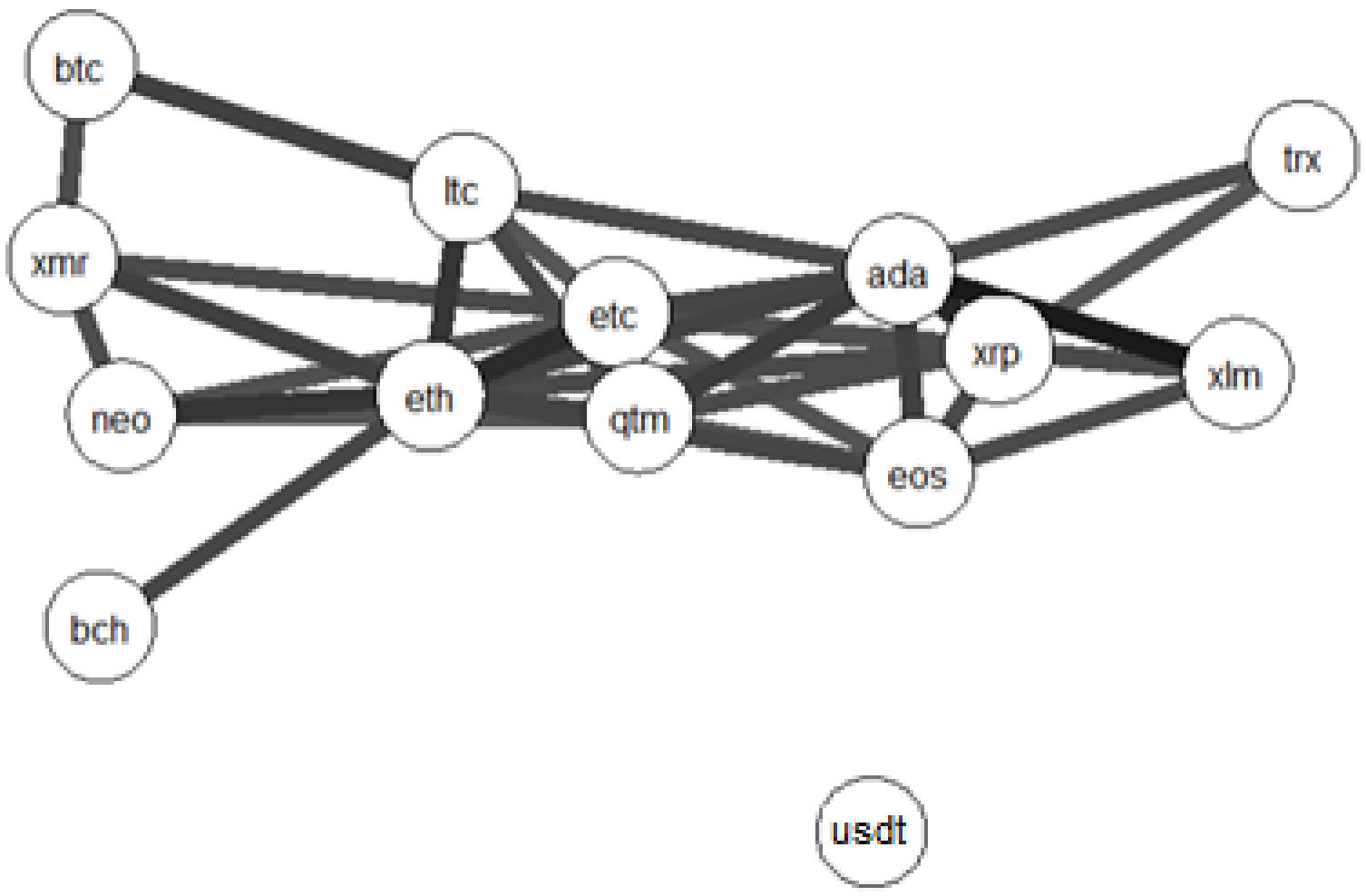

**c) Top 10 correlations**

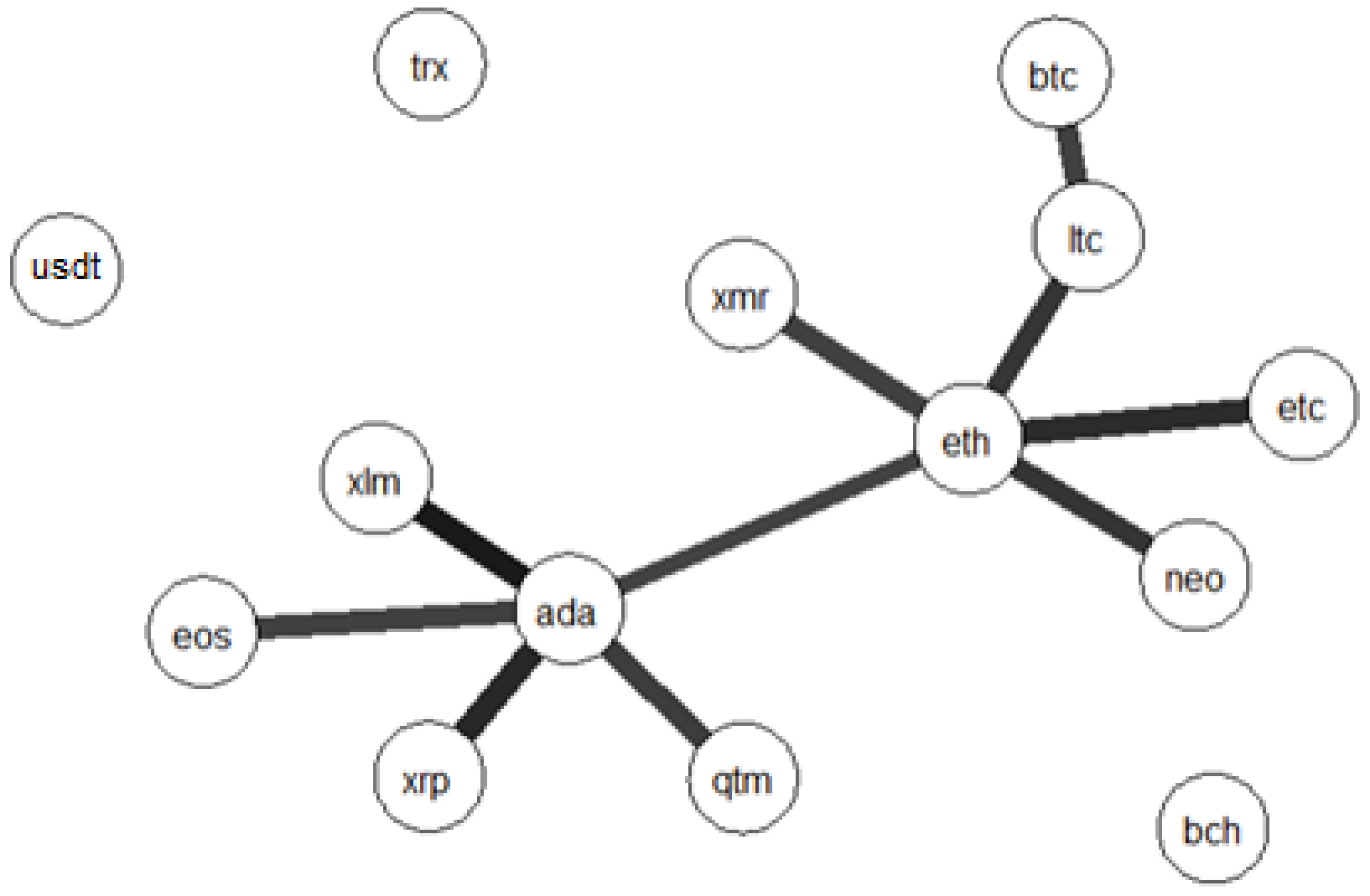

**d) Increasing threshold until group splits into two (at correlation 0.63)**

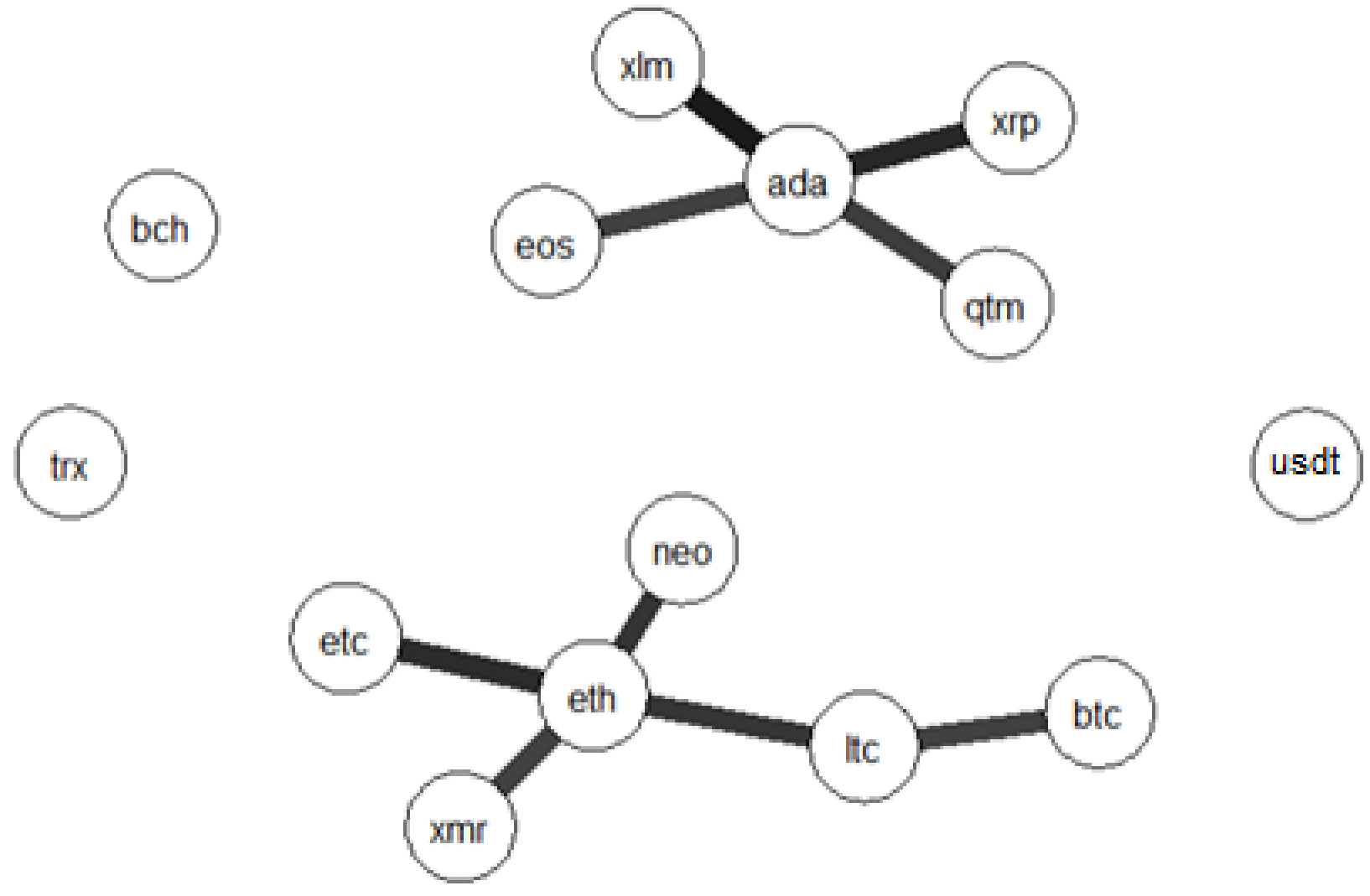

*Figure 4: Correlation Network diagrams for all cryptocurrencies; applying three different threshold methodologies. See Table 1 for explanation of abbreviations.*

Correlation network diagrams were also constructed using KT. The results were similar except: the link between xmr and eth was lost upon thresholding; qtm's correlations with eth and ada were both in the top ten; and, when the cryptocurrencies were split into two groups, eth remained linked to qtm and ada lost its connection with eos and qtm.

### 3.3.1.2 Correlation Values

| RANK | PAIR | | SR | RANK | PAIR | | SR | RANK | PAIR | | SR |
|---|---|---|---|---|---|---|---|---|---|---|---|
| **1** | ada | xlm | 0.7644 | **26** | neo | xmr | 0.5906 | **51** | bch | qtm | 0.5041 |
| **2** | ada | xrp | 0.7184 | **27** | etc | xlm | 0.5887 | **52** | xlm | ltc | 0.5026 |
| **3** | etc | eth | 0.7032 | **28** | neo | etc | 0.5807 | **53** | eos | btc | 0.5010 |
| **4** | neo | eth | 0.6752 | **29** | eos | etc | 0.5719 | **54** | eth | xlm | 0.5009 |
| **5** | eth | ltc | 0.6710 | **30** | qtm | neo | 0.5687 | **55** | xmr | xrp | 0.4934 |
| **6** | ada | qtm | 0.6468 | **31** | eos | qtm | 0.5674 | **56** | xrp | ltc | 0.4908 |
| **7** | ltc | btc | 0.6346 | **32** | trx | xrp | 0.5626 | **57** | bch | neo | 0.4893 |
| **8** | ada | eos | 0.6332 | **33** | qtm | etc | 0.5619 | **58** | eth | btc | 0.4811 |
| **9** | eth | xmr | 0.6326 | **34** | etc | ltc | 0.5609 | **59** | neo | xrp | 0.4802 |
| **10** | ada | eth | 0.6283 | **35** | ada | btc | 0.5466 | **60** | qtm | xmr | 0.4688 |
| **11** | qtm | eth | 0.6253 | **36** | xlm | btc | 0.5454 | **61** | trx | eth | 0.4683 |
| **12** | ada | etc | 0.6168 | **37** | bch | xmr | 0.5397 | **62** | eos | neo | 0.4549 |
| **13** | bch | eth | 0.6140 | **38** | qtm | xlm | 0.5388 | **63** | bch | ltc | 0.4519 |
| **14** | eos | eth | 0.6127 | **39** | eos | xmr | 0.5374 | **64** | trx | qtm | 0.4500 |
| **15** | qtm | ltc | 0.6089 | **40** | ada | neo | 0.5361 | **65** | trx | etc | 0.4455 |
| **16** | ada | ltc | 0.6086 | **41** | xmr | ltc | 0.5306 | **66** | eos | ltc | 0.4423 |
| **17** | eth | xrp | 0.6050 | **42** | trx | eos | 0.5280 | **67** | neo | ltc | 0.4327 |
| **18** | eos | xrp | 0.6008 | **43** | xlm | xmr | 0.5211 | **68** | neo | btc | 0.4288 |
| **19** | etc | xmr | 0.6002 | **44** | ada | xmr | 0.5208 | **69** | bch | eos | 0.4202 |
| **20** | xlm | xrp | 0.5998 | **45** | qtm | btc | 0.5207 | **70** | etc | btc | 0.4178 |
| **21** | qtm | xrp | 0.5994 | **46** | bch | etc | 0.5153 | **71** | xrp | btc | 0.3983 |
| **22** | eos | xlm | 0.5984 | **47** | trx | btc | 0.5138 | **72** | trx | ltc | 0.3979 |
| **23** | trx | ada | 0.5932 | **48** | trx | xmr | 0.5131 | **73** | ada | bch | 0.3674 |
| **24** | etc | xrp | 0.5925 | **49** | trx | xlm | 0.5129 | **74** | bch | xrp | 0.3571 |
| **25** | xmr | btc | 0.5912 | **50** | neo | xlm | 0.5090 | **75** | trx | bch | 0.3512 |

| RANK | PAIR | | SR |
|---|---|---|---|
| **76** | bch | btc | 0.3297 |
| **77** | trx | neo | 0.3238 |
| **78** | bch | xlm | 0.2610 |
| **79** | usdt | ltc | -0.1125 |
| **80** | bch | usdt | -0.1155 |
| **81** | qtm | usdt | -0.1198 |
| **82** | usdt | btc | -0.1265 |
| **83** | usdt | xlm | -0.1287 |
| **84** | neo | usdt | -0.1485 |
| **85** | eth | usdt | -0.1797 |
| **86** | usdt | xrp | -0.1841 |
| **87** | trx | usdt | -0.1878 |
| **88** | ada | usdt | -0.1897 |
| **89** | usdt | xmr | -0.2186 |
| **90** | eos | usdt | -0.2227 |
| **91** | etc | usdt | -0.2585 |

*Table 2: The correlations between the returns for each pair of cryptocurrencies, where correlation is measured as Spearman's rho (SR). This is for all cryptocurrencies. See Table 1 for abbreviations used.*

#### *3.3.1.3 Tests for Statistical Significance*

Two-sided Spearman's rho tests were run to evaluate the statistical significance of the different correlation values.

Excluding usdt, the p-values were less than 1% for all cryptocurrency pairs, suggesting sufficient evidence to reject a null hypothesis of no association between the cryptocurrency returns.

The cryptocurrency usdt had a p-value greater than 5% for most cryptocurrency pairs, consistent with a null hypothesis of no association between the returns. The exceptions were xrp, xmr, eos, ada and trx (p-values less than 5%) and etc (p-value less than 1%).

Kendall's tau tests corroborated the Spearman's rho tests. The only exception was the relationship between eth and usdt. This correlation was significant at 5% in Kendall's tau test, but only at 10% in Spearman's rho test. This was due to a small change in the p-value (from 5.47% to 4.47%).

### 3.3.2 Considering the Subset of Cryptocurrencies

#### *3.3.2.1 Correlation Network Diagrams*

**a) All correlations**

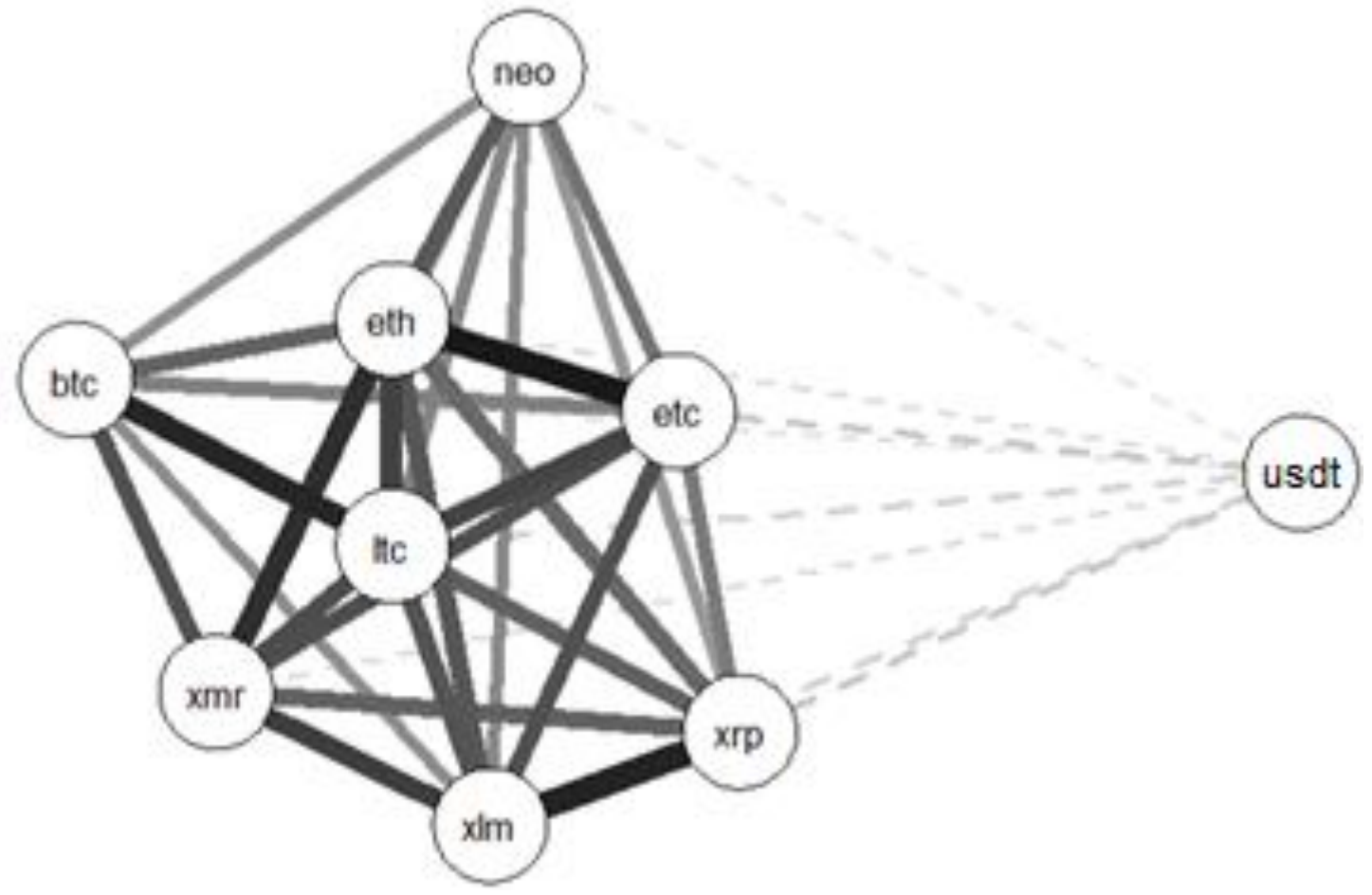

**b) Threshold where correlations jump in value (at 0.5)**

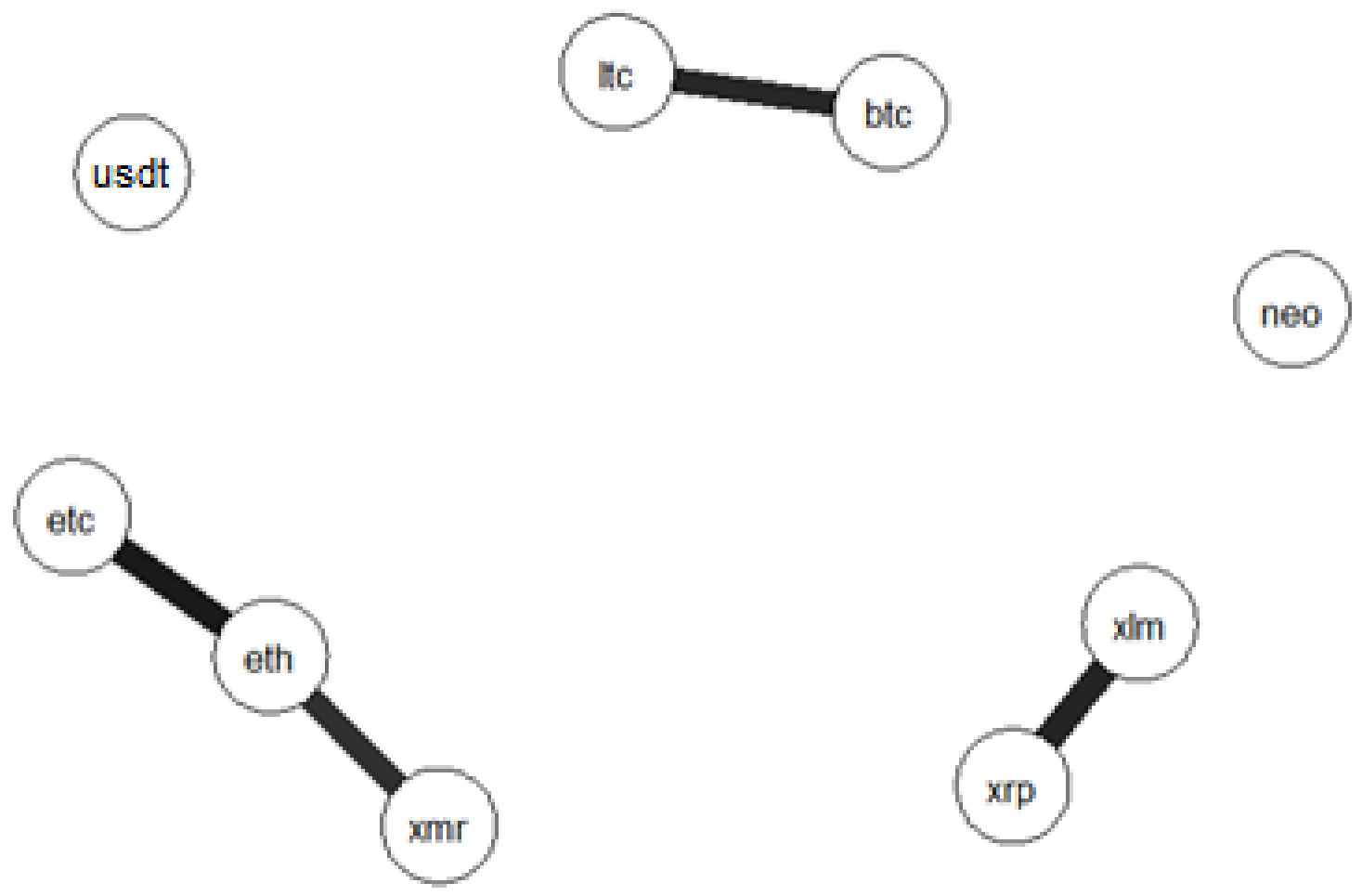

**c) Top 10 correlations**

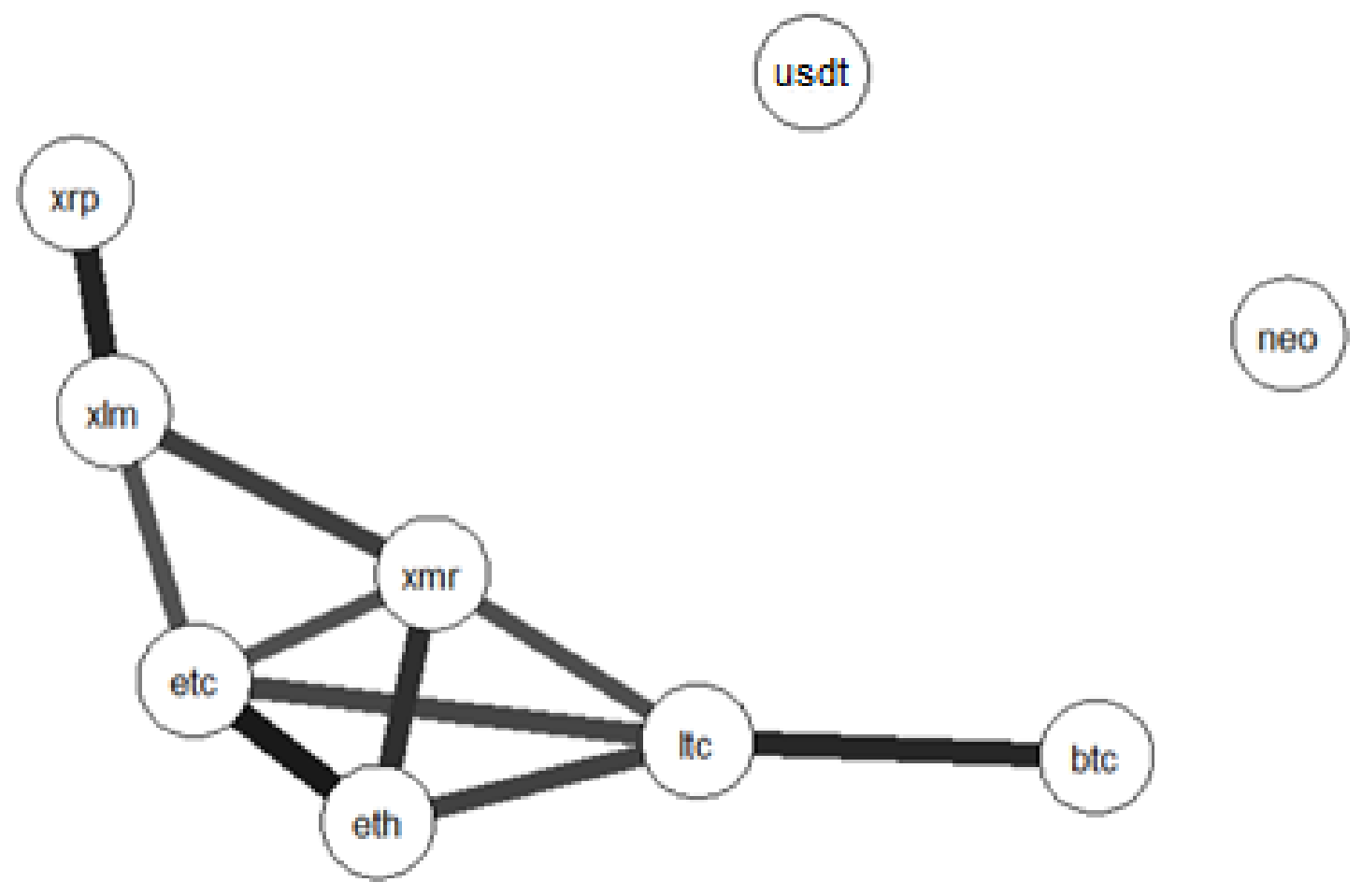

**d) Increasing threshold until group splits into two (at correlation 0.47)**

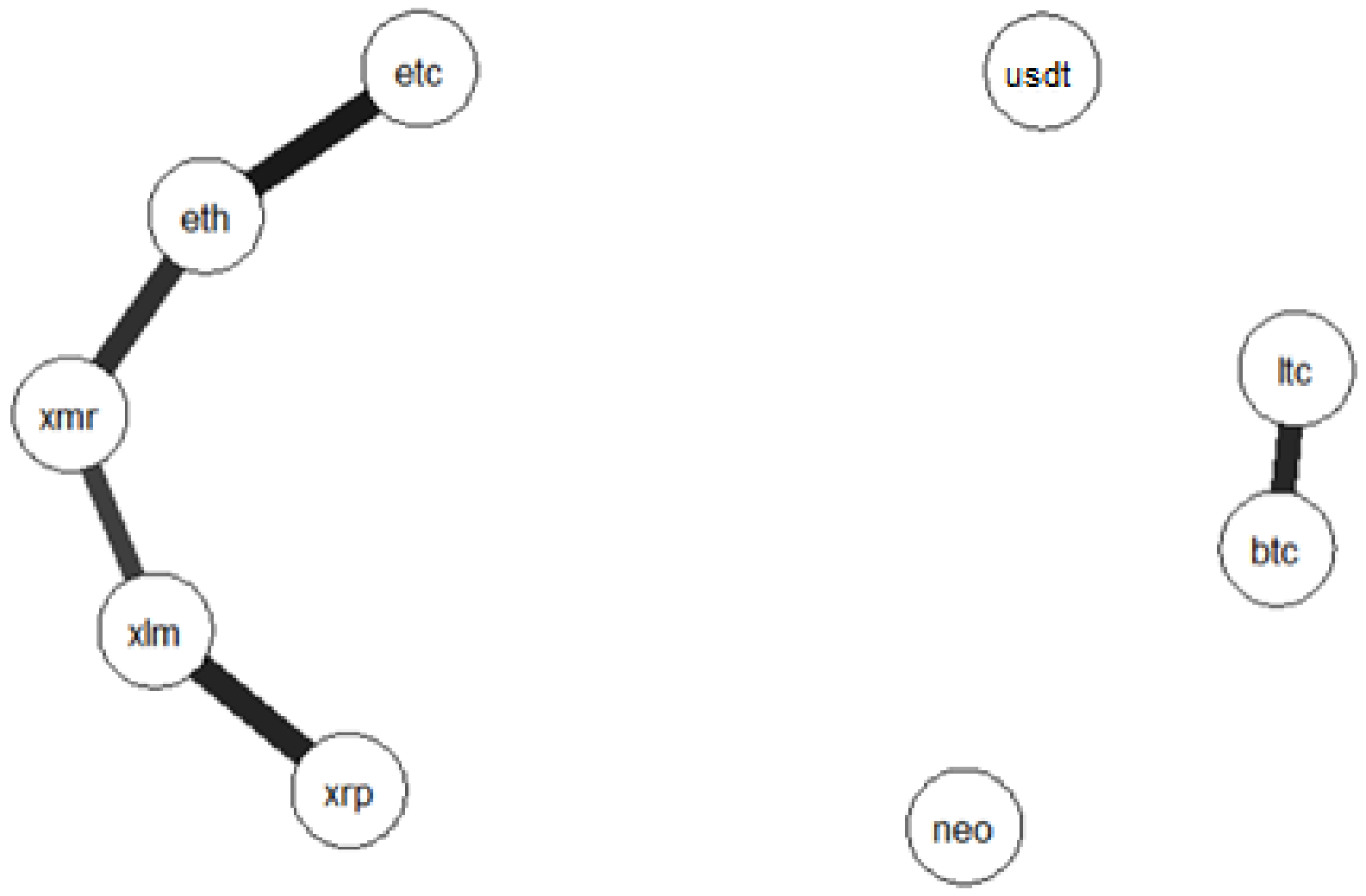

*Figure 5: Correlation Network diagrams for a subset of nine cryptocurrencies (eth, etc, usdt, ltc, btc, neo, xrp, xlm and xmr) where the data began 9 September 2016 rather than 9 November 2017. Three different threshold methodologies were applied. See Table 1 for explanation of abbreviations.*

Correlation network diagrams were also constructed using KT. The results were similar except: btc's correlation with xmr was in the top ten whilst etc's association with xmr was not; when splitting into two groups, eth was linked with ltc and xmr was not linked with xlm.

#### 3.3.2.2 Correlation Values

| RANK | PAIR | | SR | RANK | PAIR | | SR |
|---|---|---|---|---|---|---|---|
| **1** | eth | etc | 0.5680 | **19** | neo | etc | 0.3414 |
| **2** | xrp | xlm | 0.5433 | **20** | etc | xrp | 0.3241 |
| **3** | ltc | btc | 0.5356 | **21** | etc | btc | 0.3134 |
| **4** | eth | xmr | 0.5135 | **22** | neo | ltc | 0.3009 |
| **5** | xmr | xlm | 0.4753 | **23** | neo | xmr | 0.2940 |
| **6** | eth | ltc | 0.4688 | **24** | neo | xlm | 0.2883 |
| **7** | etc | ltc | 0.4577 | **25** | xlm | btc | 0.2794 |
| **8** | xmr | ltc | 0.4385 | **26** | neo | btc | 0.2754 |
| **9** | etc | xmr | 0.4345 | **27** | xrp | btc | 0.2551 |
| **10** | etc | xlm | 0.4295 | **28** | neo | xrp | 0.2530 |
| **11** | xmr | btc | 0.4255 | **29** | usdt | neo | -0.0642 |
| **12** | xlm | ltc | 0.4216 | **30** | usdt | btc | -0.0878 |
| **13** | eth | xlm | 0.4140 | **31** | usdt | xmr | -0.0996 |
| **14** | neo | eth | 0.4080 | **32** | usdt | eth | -0.1002 |
| **15** | xrp | ltc | 0.3886 | **33** | usdt | ltc | -0.1116 |
| **16** | xrp | xmr | 0.3877 | **34** | usdt | xlm | -0.1128 |
| **17** | eth | btc | 0.3849 | **35** | usdt | etc | -0.1434 |
| **18** | eth | xrp | 0.3688 | **36** | usdt | xrp | -0.1451 |

*Table 3: The correlations between the returns for each pair of cryptocurrencies, where correlation is measured as Spearman's rho (SR). This is for the subset of cryptocurrencies. See Table 1 for abbreviations used.*

#### 3.3.2.3 Tests for Statistical Significance

The presence of tied ranks prevented cor.test from calculating exact p-values; asymptotic p-values had to be used instead. Both Spearman's rho tests and Kendall's tau tests corroborated. The associations between the cryptocurrencies were all statistically significant when excluding usdt. The association between usdt and the other cryptocurrencies tended to be statistically insignificant; the exceptions being ltc, xlm, xrp and etc (at 1% significance level), and xmr and eth (at 5% significance level).

## 4. Discussion

Except for USD Tether, cryptocurrency returns were positively correlated with each other, and this association was statistically significant, providing evidence against all cryptocurrencies being unrelated and distinct (Sections 3.3.1.3 and 3.3.2.3).

The pairs Ethereum and Ethereum Classic, and Ripple and Stellar, and Bitcoin and Litecoin had the highest Spearman's rho correlation values when considering the longer dataset (Table 3).

The positive associations between Bitcoin and Litecoin, and Ethereum and Ethereum Classic were particularly robust. These correlation values remained independent of the thresholding technique applied, the time-period examined and the correlation metric used. The strong association between Bitcoin and Litecoin corroborated previous results that analysed an earlier time period from 23 June 2014 to the end of September 2016 (Osterrieder et al., 2017).

A relatively strong positive association between Ripple and Stellar was also evident in the shorter dataset beginning from 9 November 2017, indicated by their proximity in the correlation network diagrams (Figure 4 a and b). This relationship did not always remain after applying thresholding (Figure 4 c and d).

These three specific pairs can be distinguished from the other cryptocurrency pairs as follows: one cryptocurrency in the pair is a fork of the other's codebase; one cryptocurrency was launched two years after the other; and both were established before 2015. Litecoin is a fork of Bitcoin (Moskov, 2017); Ethereum Classic is a fork of Ethereum (The Ethereum Classic Community, 2016); and Stellar is a fork of Ripple (Ripple, 2017). This suggested that the cryptocurrencies that were similar in their codebases were particularly positively correlated.

An exception was Bitcoin Cash, which, despite being a fork of Bitcoin (Bitcoin Cash, 2018), was not strongly correlated to Bitcoin compared with other cryptocurrencies. This may be because Bitcoin Cash was launched in 2017, almost a decade after Bitcoin, and so its userbase is less mature.

USD Tether was an outlier, being negatively (albeit weakly) correlated with the other cryptocurrencies. A weak association was expected because Tether, unlike the other cryptocurrencies, is pegged at 1 USD Tether equal to 1 USD. However, this correlation was consistently negative in value and, at times, statistically significant. Examining trading exchange data from coingecko.com provided a possible explanation. Cryptocurrencies are often bought using USD Tether, and so USD Tether is often sold whilst a cryptocurrency is being bought. This suggests that sudden increases in the demand for cryptocurrencies (raising their prices) is likely to coincide with sudden increases in the supply of USD Tether (decreasing the USD Tether price), which could explain the observed negative correlation.

To understand the particular aspects of two cryptocurrencies' similarity that drive a similar evolution in prices over time, the correlation data was analysed to search for robust groupings. These groupings were then compared with the proposed classifications (Section 3.2).

The correlation data suggested the presence of three groups: cryptocurrencies correlated with Ethereum, cryptocurrencies correlated with Cardano and cryptocurrencies that were not particularly correlated with either cryptocurrency compared with the others. The Spearman's rho correlations between Stellar and Cardano, and Ripple and Cardano were the highest and second highest values, whilst Ethereum's relationship with Ethereum Classic, NEO and Litecoin were respectively the third, fourth and fifth highest values (Table 2). This was reflected in the correlation network diagrams (Figure 4 c and d), a result robust to changing the correlation metric to Kendall's tau.

Altering the time-period considered (Section 3.3.2) did not deteriorate Ethereum's relatively strong association with Ethereum Classic (the highest Spearman's rho value, Table 3), and Litecoin's correlation with Ethereum remained the sixth largest value. Although NEO's highest correlation remained with Ethereum, this value fell relative to the other cryptocurrency pairs (Table 3).

Comparing these groups with the proposed classifications suggested that commonalities in the token creation mechanism and the function of the token were not dominant factors in explaining why cryptocurrency prices move together. Litecoin's and Ethereum Classic's supply rises up to a cap, whilst Ethereum's supply rises indefinitely. Litecoin and Stellar were designed for transactions and yet Cardano and Ethereum are hybrid cryptocurrencies (Section 3.2).

There was limited support for the importance of being a business-orientated compared with a generic cryptocurrency. The cryptocurrencies particularly related to Ethereum fall under this paper's definition of being generic, whilst those related to Cardano are business-orientated. Using Spearman's rho (Figure 4) corroborated almost entirely with such a split (the exception being Bitcoin Cash, which is not particularly correlated with Ethereum), but such a split was not robust to changing the correlation metric.

Ripple, Cardano and Stellar can further be distinguished as using mechanisms other than Proof-Of-Work to validate transactions. Using the validation mechanism to group cryptocurrencies is limited in the need to account for why NEO, which similarly does not use Proof-Of-Work, was not part of the group related to Cardano.

A noticeable characteristic of the resulting groups was that the highest correlation values within each group were between a single, central cryptocurrency and the other members. This suggested that the central member was particularly at risk of contagion from declines in the returns of other cryptocurrencies. A possible explanation for this observation is that both Cardano and Ethereum are hybrid cryptocurrencies that can be used for both transacting value and developing applications. Changes to the viability of decentralised applications are likely to affect a cryptocurrency designed for developing applications (e.g. Ethereum Classic) as well as the related hybrid (Ethereum), whilst not affecting as much a cryptocurrency intended for transactions (e.g. Litecoin). Similarly, a change impacting Litecoin is likely to be more relevant to Ethereum than Ethereum Classic.

Although this paper provides evidence to support similar cryptocurrencies as being positively associated, it cannot prove which aspects of similarity are important, beyond suggesting that the token's functionality and supply mechanism were unlikely to be the dominant features.

This paper can also not prove whether it is the actual similarity or the perceived similarity between cryptocurrencies that is important. Investors are likely to perceive that if a cryptocurrency is forked from another that these two cryptocurrencies are similar. Hence, if investors see that a parent cryptocurrency's price has fallen (e.g. Ethereum), they might then assume that similar cryptocurrencies are also likely to fall (e.g. Ethereum Classic), causing them to sell the child cryptocurrency, leading to a reduction in its price. This may occur even if there are material differences between the two cryptocurrencies.

This alternative explanation is limited by the results observed for Bitcoin Cash. Bitcoin Cash both shares Bitcoin's name and is a fork from Bitcoin's codebase, and so psychologically investors are likely to view the two cryptocurrencies as similar. However, the dataset revealed that these two cryptocurrencies were not particularly correlated.

## 5. Conclusion

This paper has found that the correlation between cryptocurrencies is particularly strong and robust for cryptocurrencies where one is a fork of the other, with the notable exception of Bitcoin and Bitcoin Cash. This suggests that, although speculation is important (Bouoiyour & Selmi, 2015; Kristoufek, 2013), there may be other features of cryptocurrencies that shape the evolution of prices over time, and thus which are relevant to valuing cryptocurrencies.

Which aspects of similarity are important in shaping the change in prices over time remains unresolved. However, this paper does provide evidence to suggest that how tokens are created and their functionality are not particularly important determinants. This suggests that investors should look beyond the simpler characteristics of a cryptocurrency's token to understand how the value of the token will evolve over time. This could include evaluating the entities upon which a cryptocurrency system depends to succeed (Burnie et al., 2018).

This paper has further revealed the value of correlation networks as a tool for visually exploring whether intuitions held about the interrelationships between financial assets (here cryptocurrencies) are reflected in the data. Whilst regression analyses often assume linearity when comparing different financial variables (Ciaian et al., 2016, 2018; Kristoufek, 2013), switching to Spearman's rho or Kendall's tau enables a broader category of associations to be evaluated.

Future work could investigate further candidates for value-relevant, time-invariant cryptocurrency characteristics, a wider sample of even more heterogeneous cryptocurrencies and could extend the dataset as more data becomes available.

## Appendix

The below provides links to the different websites and whitepapers used to substantiate the proposed groupings (Section 3.2).

Bitcoin

- Whitepaper: https://bitcoin.org/bitcoin.pdf
- Omni-layer protocol, giving Bitcoin some functionality in developing applications: http://www.omnilayer.org/

Litecoin

- Wiki: https://litecoin.info/index.php/Main_Page
- Websites: https://litecoin.org/ and https://litecoin.com/

Ethereum

- Whitepaper: https://github.com/ethereum/wiki/wiki/White-Paper
- Ethereum is used for buying goods/services, as discussed in the following web-link: https://www.finder.com/ethereum-classic

Ethereum Classic

- Website: https://ethereumclassic.github.io/
- Monetary policy: https://www.etcdevteam.com/blog/articles/a-joint-statement-ecip1017.html
- Ethereum Classic is rarely used for buying goods/services, as discussed in the following web-link: https://www.finder.com/ethereum-classic

Monero

- Website: https://getmonero.org/
- Detailed comparison with Bitcoin: https://www.monero.how/why-monero-vs-bitcoin

NEO

- Documentation: http://docs.neo.org/en-us/

Bitcoin Cash

- Website: https://www.bitcoincash.org/

Tron

- Archived Whitepaper: https://ipfs.io/ipfs/QmWh3LEWUQN8LsoHerQecmwfACXAPNKE9wigx6t9dLitmE/tron/Tron-Whitepaper-1031-V18-EN.pdf
- Evidence of Tron as an Ether token available from Etherscan: https://etherscan.io/token/Tronix

Cardano

- Whitepaper: https://whycardano.com/
- Discussion on Cardano website detailing how ada (the Cardano token) can be used for buying goods and services: https://www.cardanohub.org/en/shop-with-cardano/

Qtum

Two whitepapers:

- Dai, P., Mahi, N., Earls, J., & Norta, A. (2017). Smart-Contract Value-Transfer Protocols on a Distributed Mobile Application Platform. No Publisher. Retrieved from https://qtum.org/uploads/files/a2772efe4dc8ed1100319c6480195fb1.pdf
- Qtum Foundation. (2017). Qtum Blockchain Economy Whitepaper. Qtum Foundation. Retrieved from https://qtum.org/uploads/files/ef2723f33deef1875ef17361f7c696ef.pdf
- Few merchants accept Qtum; discussed in following web-link: https://www.finder.com/uk/qtum

Ripple

- Website: https://ripple.com/
- Wiki: https://wiki.ripple.com
- GitHub repo: https://github.com/ripple/rippled
- Details on how transaction costs reduce the amount of Ripple over time: https://ripple.com/build/transaction-cost/
- Details specific to XRP, Ripple’s token: https://ripple.com/xrp/
- Details on how those validating transactions are chosen: https://ripple.com/technical-faq-xrp-ledger/

EOS

- Whitepaper: https://github.com/EOSIO/Documentation/blob/master/TechnicalWhitePaper.md

Stellar

- Basics of Stellar as a network to connect financial products with customers: https://www.stellar.org/how-it-works/stellar-basics/
- Discussion of Stellar Lumens, Stellar's token and how it is distributed: https://www.stellar.org/lumens/
- Details of Federated Byzantine Agreement: https://www.stellar.org/papers/stellar-consensus-protocol.pdf

Tether
- Whitepaper: https://tether.to/wp-content/uploads/2016/06/TetherWhitePaper.pdf

The above links were all last accessed on 10 March 2018.

## Acknowledgements

This work was supported by The Alan Turing Institute under the EPSRC grant EP/N510129/1 and Turing award number TU/C/000028.